\documentclass{bmvc2k}
\usepackage{graphicx}
\usepackage{amsmath}
\usepackage{amssymb}
\usepackage{booktabs}
\usepackage{microtype}
\usepackage{adjustbox}
\usepackage{listings}
\usepackage{pythonhighlight}
\usepackage{float}
\usepackage{tablefootnote}
\usepackage{tabularx}
\usepackage{wrapfig}
\usepackage{duckuments}
\usepackage{caption}
\usepackage{subcaption}
\usepackage[T1]{fontenc}
\usepackage{multirow}

\usepackage[capitalise,nameinlink,noabbrev]{cleveref}
\newcommand{\etal}{\textit{et al}.}
\newcommand{\ie}{\textit{i}.\textit{e}.}

\title{Content and Style Aware Audio-Driven Facial Animation}

\addauthor{Qingju Liu}{qingju.liu@flawlessai.com}{1}
\addauthor{Hyeongwoo Kim}{hyeongwoo.kim@imperial.ac.uk}{2}
\addauthor{Gaurav Bharaj}{gaurav.bharaj@gmail.com}{1}

\addinstitution{
Flawless AI, UK
}
\addinstitution{
Imperial College London, UK
}

\runninghead{LIU et al}{Content and Style Aware Audio-Driven Facial Animation}

\def\etal{\emph{et al}\bmvaOneDot}

\begin{document}

\maketitle

\begin{abstract}
Audio-driven 3D facial animation has several virtual humans applications for content creation and editing. While several existing methods provide solutions for speech-driven animation, precise control over \emph{content} (what) and \emph{style} (how) of the final performance is still challenging.
We propose a novel approach that takes as input an audio, and the corresponding text to extract temporally-aligned content and disentangled style representations, in order to provide controls over 3D facial animation. Our method is trained in two stages, that evolves from audio prominent styles (how it sounds) to visual prominent styles (how it looks).
We leverage a high-resource audio dataset in stage I to learn styles that control speech generation in a self-supervised learning framework, and then fine-tune this model with low-resource audio/3D mesh pairs in stage II to control 3D vertex generation. We employ a non-autoregressive \emph{seq2seq} formulation to model sentence-level dependencies, and better mouth articulations.
Our method provides flexibility that the style of a reference audio and the content of a source audio can be combined to enable audio style transfer. Similarly, the content can be modified, e.g. muting or swapping words, that enables style-preserving content editing.
\end{abstract}

%-------------------------------------------------------------------------
%%%%%%%%% BODY TEXT
\section{Introduction}
\label{sec:intro}

Virtual human applications, such as, AR/VR, CGI, and telepresence has led to an increased focus on modeling audio-driven 3D facial animation. We desire a system that, given an unseen audio, provides precise control over the generated facial animation, where the articulated \emph{content} is disentangled from \emph{style} (attributes such as identity, emotions, idiosyncrasies). Such flexible control offers the opportunity to change content and the visual style with mix 'n match, while alleviating the need to record new performances. %For example, \emph{style} from an audio can be used with \emph{content} from another audio seamlessly. %animation system can then be used to render 3D human faces and downstream tasks such as, controlled photorealistic facial image rendering, emotion editing, and audio-driven performance transfer.
Existing audio-driven facial animation systems~\cite{lewis1986automated, brand1999voice, edwards2016jali,cudeiro2019capture, thies2020neural, richard2021meshtalk, fan2022faceformer, ma2024diffspeaker, peng2023emotalk} don't provide explicit \emph{content} and \emph{style} representations that we can manipulate separately, where source audio is often mapped to either simple chart tables~\cite{lewis1986automated, edwards2016jali} or a latent embedding space that can't be interpreted semantically \cite{richard2021meshtalk, fan2022faceformer}. While explicit conditioning factors such as one-hot identity can be integrated \cite{cudeiro2019capture, fan2022faceformer, ma2024diffspeaker, peng2023emotalk}, however, such a representation has limited capacity to model rich and highly diverse styles.

\begin{figure*}
  \centering
\includegraphics[width=\linewidth]{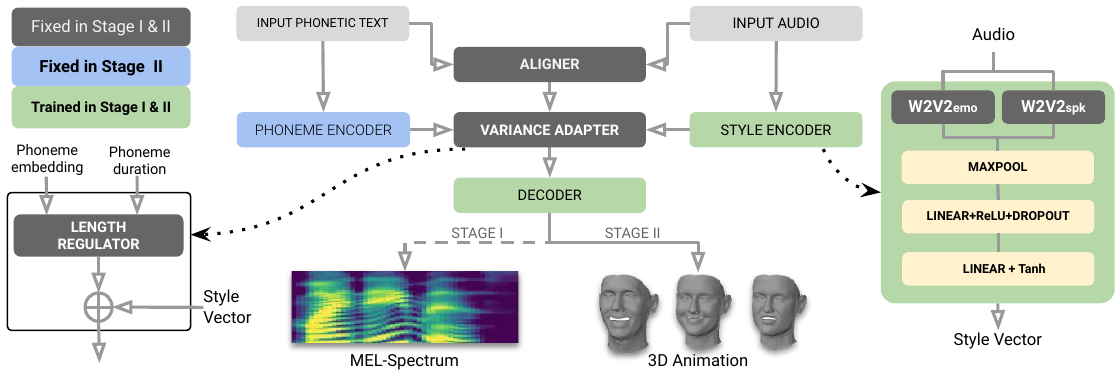}
    \caption{ \textbf{System Overview.} During Stage I, styles are learnt via Mel-spectrum reconstruction. Stage II fine-tunes the network for the highly correlated audio-driven animation task. Left shows the architectural details of the variance adapter, and right the style encoder that outputs disentangled styles, initialised with two pre-trained Wav2Vec2~\cite{baevski2020wav2vec} networks.}
  \label{fig:overview}
\end{figure*}

While facial audio-video dataset are available in abundance~\cite{wang2020mead, zhu2022celebvhq}, large high-quality audio-3D mesh datasets are costly to create, and require post-processing for consistent mesh topology~\cite{cudeiro2019capture}. This data limitation has constrained the effective generalized learning for audio-driven facial animation. 
% \gb{Recent advances in generative methods has witnessed high potential of knowledge transfer for correlated fields\tocite}. For example, 
Text-to-Speech (TTS)~\cite{tocotron2, kumar2019melgan,ren2020fastspeech} is a highly-correlated task, where high-fidelity speech can be synthesized preserving a speaker's characteristics -- timbrel, prosody, accent, etc. Style controlled TTS~\cite{sun2020prosodymodel,huang2022generspeech, Li2022multiscale,li2022styletts} is most similar to our use-case. The audio speech and the corresponding visual speech \ie, talking head videos~\cite{fried2019text,thies2020neural} are highly correlated since both modalities are generated by the articulation system. For instance, the work in~\cite{yu2019durian} produces speech and facial expression simultaneously. Thus, knowledge in \emph{high-resource} TTS methods can be potentially transferred to \emph{low-resource} audio-driven facial animation. To investigate this, we present a novel two stage approach with knowledge transfer for audio-driven facial animation, see~\cref{fig:overview}.

In Stage I, we train an (audio+text to Mel-spectrum) model for the speech synthesis (TTS) task via spectrum reconstruction. Exploiting spectrum reconstruction to learn styles has also been used in~\cite{li2022styletts, min2021meta}, where the style encoders are learned from scratch. To encourage disentanglement of audio styles, we initialised the style encoder with two sub-networks pre-trained for speaker verification and emotion classification respectively.

In Stage II, we train an (audio+text to mesh) model to learn \emph{style-aware} audio-driven facial animation, where the network structure is modified over the top output layers (vs. Stage I), see~\cref{sec:method}. Stage II essentially fine-tunes the Stage-I network, and alleviates the need for a larger audio-mesh dataset. % for the highly-correlated audio-driven animation task. 
Driven by tasks of different modalities, the learned styles evolve from audio-prominent (Mel-spectrum) to visual prominent (3D meshes). To the best of our knowledge, we are the first to attempt such a modality-evolving formulation. 

We summarize our contributions below: (1) A novel formulation for \emph{content} and \emph{style} aware audio-driven facial animation where disentangled styles evolve from adapting audio speech self-reconstruction to facial animation generation. (2) Non-autoregressive (NAR) modeling for audio and text pair with a modified computational-efficient Laplacian loss. (3) Our formulation leads to both audio style transfer and content modification.

% https://docs.google.com/drawings/d/16VScLKJCsmm4Us496uv-t-HiCTDpjKBvApN5liFnDjw/edit

\section{Related Works}
\label{sec:relatedworks}
% \paragraph{Audio-driven Mesh Vertex Animation.} 
\textbf{Audio-driven Mesh Vertex Animation.} Several audio-driven facial animation methods generate animated 3D meshes directly \cite{karras2017audio, cudeiro2019capture,richard2021meshtalk, fan2022faceformer, yang2023semi, ma2024diffspeaker} from audio. The early work from Kerras~\etal\cite{karras2017audio} builds a subject dependent regression model with a learned latent emotional state. Cudeiro~\etal\cite{cudeiro2019capture} takes DeepSpeech~\cite{hannun2014deep} features and one-hot speaker embedding to regress the 3D mesh in FLAME~\cite{li2017learning} 3DMM. 
Meshtalk~\cite{richard2021meshtalk} presents a two-stage method with semantically-disentangled expressive categorical hidden vectors and an auto-regressive (AR) network to sample from this space to control the mesh decoder. 

FaceFormer~\cite{fan2022faceformer} builds a multi-subject pipeline, featuring a Wav2Vec2~\cite{baevski2020wav2vec} encoder, and an AR transformer decoder with biased attention. Several other works \cite{xing2023codetalker,thambiraja2022imitator} extend FaceFormer. For instance, CodeTalker~\cite{xing2023codetalker} uses a discrete codebook of motion primitives learnt via vector quantized autoencoder (VQ-VAE), to avoid the regression-to-mean problem that leads to over-smoothed facial motions. Imitator \cite{thambiraja2022imitator} allows adaption to new users. 

The method in~\cite{Aylagas2022} uses a conditional variational autoencoder (cVAE) conditioned on contextual speech features. A diffusion-based transformer backbone is employed in \cite{ma2024diffspeaker} with a biased
conditional attention mechanism. The work in \cite{yang2023semi} uses facial images to bridge the audio and 3D mesh, where the visual encoder is essentially a 3D face tracker. 

\noindent\textbf{Audio-driven 3D Blendshapes/Visemes/Portraits.} Other works use animation representations based on 3D head models, e.g. blendshapes and visemes~\cite{edwards2016jali,zhou2018visemenet,medina2022speech,abdelaziz2020audiovisual,yi2020audio,peng2023emotalk}. Such methods often require high-quality data that is obtained via either 3D scans followed by 3DMM registration, or 3D face tracking on recordings in well-controlled environments. 

Although, we desire facial animation as the final output, several methods use sparse landmarks~\cite{wang2020speech, prajwal2020lip, zhou2020makelttalk, ye2023geneface}, feature maps~\cite{lu2021live}, blendshapes~\cite{thies2020neural,ji2021audio}, or vertex animation~\cite{lahiri2021lipsync3d} as intermediate representations for facial image synthesis. Works in~\cite{fried2019text,li2021write} use a text-blendshape-human head (image) formulation. Methods, such as, \cite{ji2021audio, ji2022eamm, li2021write, gururani2022space} model emotion-aware talking head image synthesis with intermediate blendshapes/edge maps. Some recent works use diffusion models \cite{stypulkowski2023diffused, shen2023difftalk} for portraits animation.

\noindent\textbf{Style Controlled Speech Synthesis.} 
\begin{figure}
\centering
\begin{subfigure}{.24\textwidth}
  \centering
  \includegraphics[width=\textwidth]{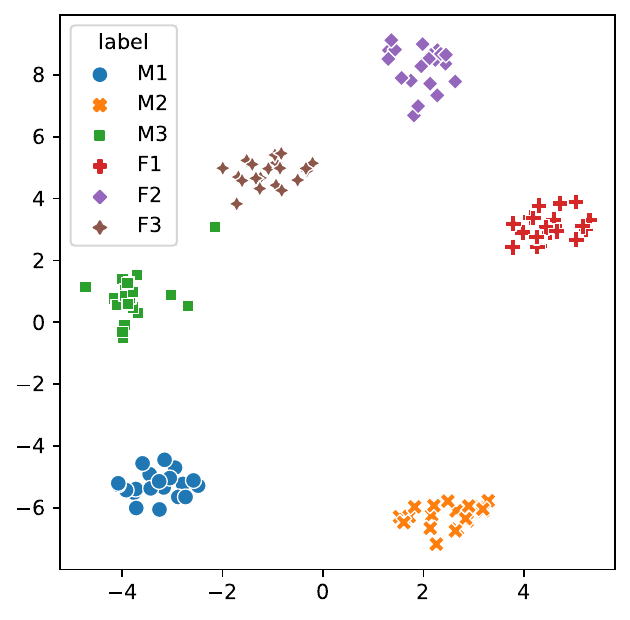}
  \caption{Subject (proposed)}
  \label{fig:esd1}
\end{subfigure}%
\begin{subfigure}{.24\textwidth}
  \centering
  \includegraphics[width=\textwidth]{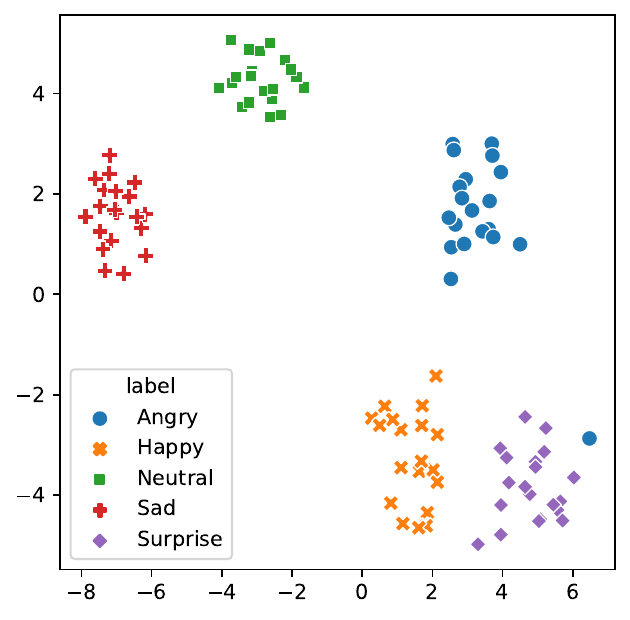}
  \caption{Emotion (proposed)}
  \label{fig:esd2}
\end{subfigure}
\begin{subfigure}{.245\textwidth}
  \centering
  \includegraphics[width=\textwidth]{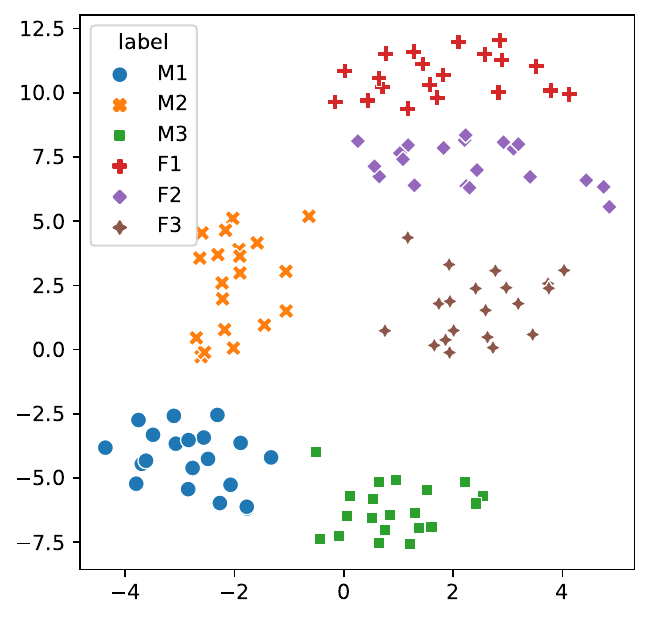}
  \caption{Subject (\cite{min2021meta})}
  \label{fig:esd_metaspeech}
\end{subfigure}
\begin{subfigure}{.24\textwidth}
  \centering
  \includegraphics[width=\textwidth]{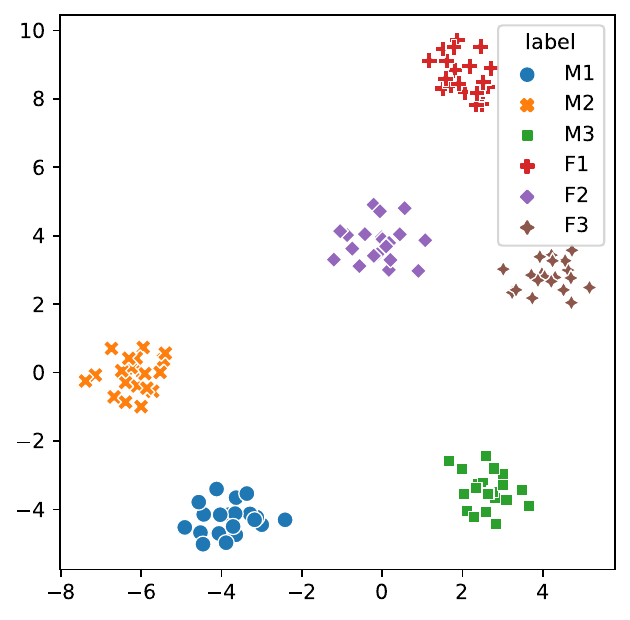}
  \caption{Subject (\cite{wan2020generalized})}
  \label{fig:esd_resemblyzer}
\end{subfigure}
\caption{The T-SNE plot of the style vectors applied onto ESD dataset after Stage I training over (a) six randomly chosen subjects, and (b) one subject with different emotions. (c) The method \cite{min2021meta} cannot cluster well speakers, while (d) \cite{wan2020generalized} does not cover other attributes.}
\label{fig:stylevector}
\end{figure}
Styles in TTS are challenging to model and lack a clear definition. Explicit labels such as emotions, or speaker codecs can be used as styles~\cite{luong2017adapting}. Styles can also be learned in an unsupervised manner. For instance, in \cite{skerry2018towards} a prosody embedding from a reference audio is learnt to synthesize speech in the Tacotron pipeline~\cite{tocotron2}. Similarly, \cite{wang2018style} proposes global style tokens (GSTs) that effectively take into account various noise and speaker factors. Li~\etal\cite{li2022styletts} employs a two stage method, where style vectors are extracted in the first stage via Mel-spectrum reconstruction.

Multi-scale hierarchical style can also be exploited. Sun~\etal\cite{sun2020prosodymodel} models prosody at utterance, word and phoneme levels with a cVAE, while Li~\etal\cite{Li2022multiscale} learns utterance-level and quasi-phoneme-level style features. GenerSpeech~\cite{huang2022generspeech} employs global speaker and emotion embeddings, together with prosody styles at frame, phoneme, and word levels.

% \vspace{-20px}

\section{Our Method}
\label{sec:method}

To develop a style-aware audio-driven facial animation method, the style needs to be disentangled from content. We propose transferring style information learned from high-resource speech data, to low-resource audio-visual data, with a two stage scheme.

\noindent\textbf{Stage I: Mel-Spectrum Reconstruction.} 
\noindent
In Stage I, we learn styles that control speech generation via Mel-spectrum reconstruction. We employ a backbone structure (phoneme encoder, variance adapter, decoder) similar to Fastspeech2~\cite{ren2020fastspeech}, \cref{fig:overview}, an end-to-end parallel speech synthesis network. The phoneme encoder and the decoder both use stacked self-attention layers \cite{vaswani2017attention}. The variance adapter is simplified, shown in bottom left of \cref{fig:overview}. The phoneme duration is calculated via forced-alignment of the driving audio with paired text using the Montreal forced aligner (MFA)\cite{mcauliffe2017montreal}.

Our formulation contains disentangled content and style, where the content is the length-regulated phoneme sequence. To encourage the disentanglement of style from content, the style encoder, shown in bottom right of \cref{fig:overview}, is designed as follows.

Resembling the two global tokens used in \cite{huang2022generspeech}, we initialise the style encoder with two pre-trained Wav2Vec2 \cite{baevski2020wav2vec} sub-networks for speaker (spk) verification and emotion (emo) classification tasks, denoted as W2V2$_{spk}$ and W2V2$_{emo}$, respectively. Since they are trained on two tasks that are not necessarily relying on content, and we use two models \cite{yang2021superb} pre-trained on large datasets with a high content diversity, therefore disentanglement are achieved. The latent hidden embeddings from the above sub-nets are then combined to obtain the final style vector. We use Mel-spectrum reconstruction loss to guide Stage I training: 
\begin{align}
    {\cal L}_{stageI} = {MAE}(a(t,f), {\hat a}(t,f)), 
\end{align}
\noindent
where $a(t,f)$ and ${\hat a}(t,f)$ is the source and predicted Mel-spectrum at the time-frequency unit $(t,f)$, and ${MAE}(\cdot, \cdot)$ represents the mean absolute error. The predicted ${\hat a}(t,f)$ is the top layer output from the decoder followed by a linear layer mapping ${\cal F}_{s1}(\cdot)$, 
\vspace{-2px}
\begin{align}
{\hat a}(t,f) = {\cal F}_{s1}(\text{Decoder}(\texttt{Text}, \bf s)),
\end{align}
\vspace{-2px}
\noindent
where $\texttt{Text}$ is the content component in the form of duration regulated phoneme embedding, and $\bf s$ denotes the extracted style vector.

\noindent\textbf{Stage II: 3D Vertex Generation.}
\noindent
We slightly modify the model architecture at Stage II. The top layer of the decoder is mapped to the high-dimensional vertex space $\mathbb{V}\in {3N_v}$ instead, where $N_v$ denotes the number of vertices in a head model topology. This mapping process ${\cal F}_{s2}(\cdot)$ is achieved by a sequential network containing a linear layer with a \texttt{Tanh} activation and another linear layer, to predict the 3D mesh ${\bf \hat V} \in {3N_v \times T}$ spanning in total $T$ frames,
\vspace{-5px}
\begin{align}
{\bf \hat V} = {\cal F}_{s2}(\text{Decoder}(\texttt{Text}, \bf s)). 
\end{align}
% \vspace{-5px}
\noindent
During Stage II training, only a few parameters are fine-tuned to adapt to the new task. Few landmark vertices, e.g. the 68 Multi-PIE \cite{gross2010multi} are perceptually and visually important. Particularly, we are more interested in the 20 key mouth landmarks that highlight the mouth motion. We directly extract them from the full vertex-set ${\bf \hat V}$, denoted as ${\bf \hat K}$. We use the following loss to train Stage II:
\vspace{-5px}
\begin{equation}
    {\cal L}_{stageII} = {\cal L}_{geometry} + {\cal L}_{temporal} + \lambda {\cal L}_{lap-mod}, 
    \label{s2loss}
\end{equation}
here ${\cal L}_{geometry}$ is the sum of MAE losses over vertices ${\bf \hat V}$ and landmarks ${\bf \hat K}$. To prioritize mouth movements that are more correlated with the audio, a weighting mask is applied onto the vertex space, where vertices near the mouth region are given higher priority. Similarly, higher weights are given to the interior mouth landmarks when calculating the landmark loss. ${\cal L}_{temporal}$ is the sum of MAE losses over the first-order difference (dynamics) of vertices and landmarks, in order to promote temporal smoothness. %\looseness=-2

${\cal L}_{lap-mod}$, weighted by $\lambda$, is the Laplacian loss that boosts spatial smoothness and mitigates high-frequency noise in the generated meshes. The Laplacian loss can be directly applied onto the predicted surface ${\bf \hat V}$. However, this involves a very high computational complexity, especially for a head topology with a large number of vertex $N_v$. The complexity scales linearly with the number of batched input frames. Besides, the computation needs back propagation through all unfrozen layers of the pipeline. Denote the last layer weight matrix in ${\cal F}_{s2}(\cdot)$ as ${\bf W} \in {3N_v \times D}$, and each frame of ${\bf \hat V}$ is essentially a linear combination of column basis in ${\bf W}$; thus we approximate ${\cal L}_{lap-mod}({\bf \hat V})$ with ${\cal L}_{lap-mod}(\bf W)$. This greatly reduces the complexity when $D$ is small, and the associated gradient back-propagation stops after only one layer, essentially forming ${\cal L}_{lap-mod}$ as a regularisation term. For Stage II training, most of the module are fixed (\cref{fig:overview}). Particularly, in order to allow learning prominent visual difference that are not confined by audio speaking styles, the style encoder is further fine-tuned, together with the decoder.\looseness=-2

\begin{figure}[t]
\centering
\begin{subfigure}{.24\textwidth}
  \centering
  \includegraphics[width=\textwidth]{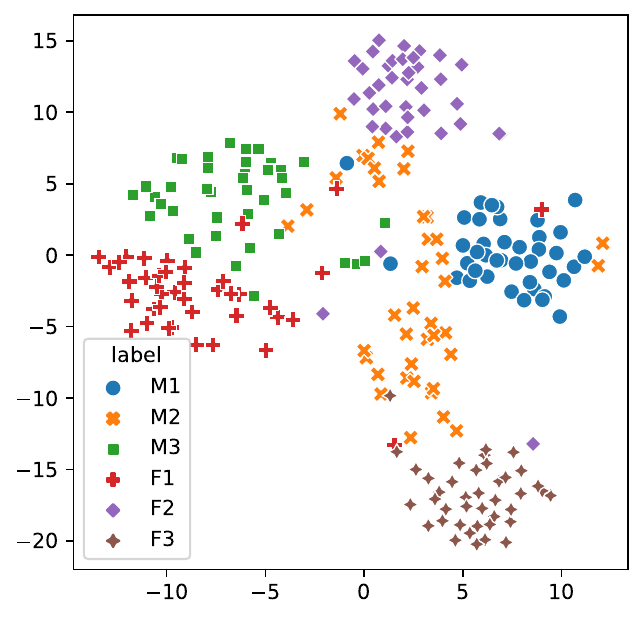}
  \caption{VOCAset (Stage I)}
\end{subfigure}%
\begin{subfigure}{.24\textwidth}
  \centering
  \includegraphics[width=\textwidth]{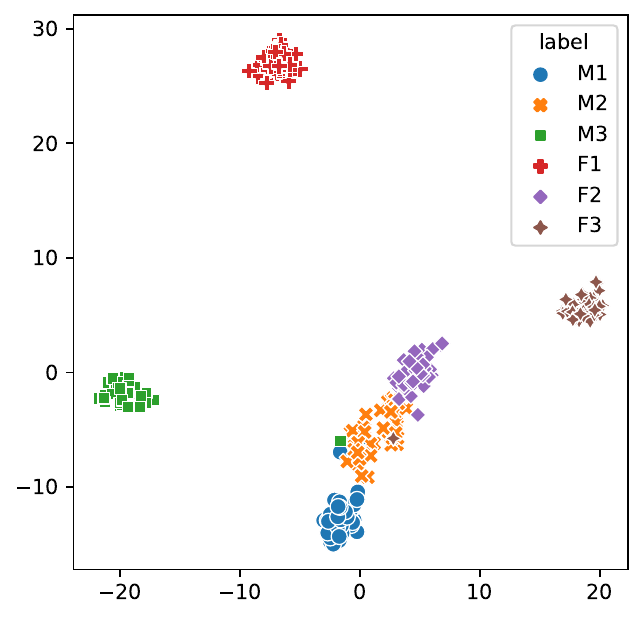}
  \caption{VOCAset (Stage II)}
\end{subfigure}
\begin{subfigure}{.245\textwidth}
  \centering
  \includegraphics[width=\textwidth]{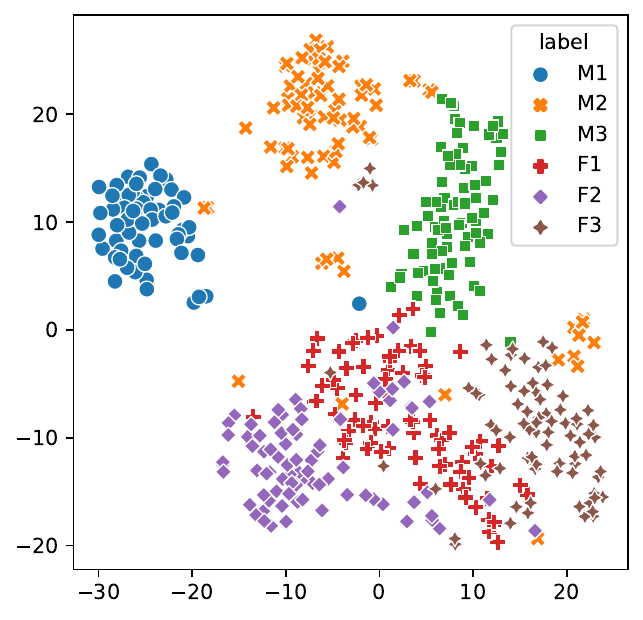}
  \caption{BIWI (Stage I)}
\end{subfigure}
\begin{subfigure}{.24\textwidth}
  \centering
  \includegraphics[width=\textwidth]{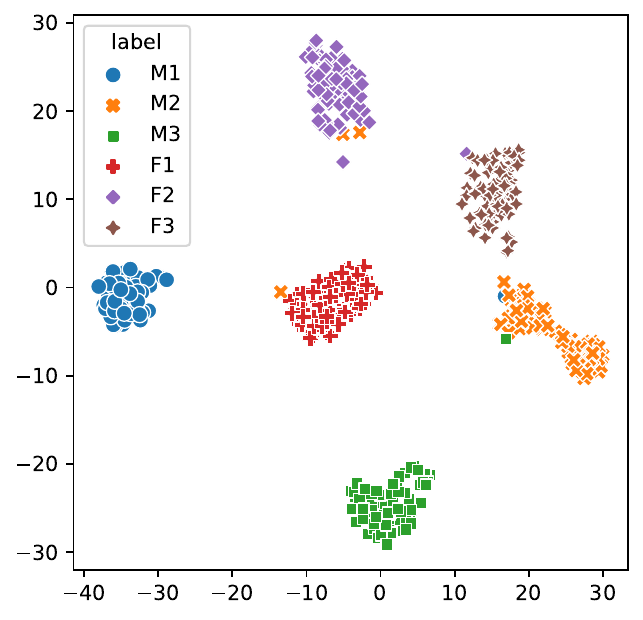}
  \caption{BIWI (Stage II)}
\end{subfigure}
\caption{The learned styles evolve from Stage I to Stage II with different modality focuses.}
\label{fig:s1s2stylevector}
\end{figure}

\noindent\textbf{Inference pipeline.} Once the model is built after the above two stage training, the inference pipeline takes in one driving audio together with the associated transcript (to provide content) and one reference audio (to provide style) in a mix ’n match manner, in order to directly generate the facial animations. During inference, Stage I model won't be employed and we only use the evolved model weights in Stage II model. 
\section{Results}
\label{sec:results}
\noindent
\textbf{Datasets.} We use two audio datasets for Stage I, emotional speech database (ESD)~\cite{zhou2022emotional} and LJSpeech~\cite{ljspeech17}. All the English data in ESD is used, denoted as ESD-EN with the total duration of around 13.5 hours. ESD-EN contains utterances from 10 speakers and 5 emotional categories. Yet, it greatly suffers from limited content coverage with only 350 parallel transcripts. On the other hand, LJSpeech covers a high diversity in content with varying transcripts for each recorded passage. To balance the two datasets, we chose 2000 sequences from LJSpeech (similar amount of sequences for each subject in ESD-EN), denoted as LJSpeech-2K, with total duration around 3.7 hours. Combination of these two datasets balances content and style coverage and avoids over-fitting to a particular speaker or limited transcripts.\looseness=-1

We use two audio-mesh datasets during Stage II to train and test of our pipeline -- the VOCAset~\cite{cudeiro2019capture} and BIWI~\cite{BIWI2010}, both contain registered 3D vertices from high quality scans. The VOCAset employs the FLAME~\cite{li2017learning} head topology with 5023 vertices per mesh, sampled at 60 fps (downsampled to 30 fps). It contains recordings from 12 subjects, each uttering 40 sequences in neutral tone with different transcripts. The BIWI dataset uses a dense topology (23370 vertices per mesh), sampled at 25 fps. It includes 14 subjects, each uttering 40 parallel sequences in neutral and affective tones.\looseness=-1
% We randomly chose 6 subjects from both datasets and plot the associated identity template and per-vertex variations in~\cref{fig:identity}. 

\noindent
\textbf{Data Preprocessing.} All audio signals are normalised with mono-channel, 16KHz sampling rate, and maximum magnitude of 1.0. A 80 dimensional Mel-spectrum is extracted with 1024-size sliding Hanning window overlapped at 256 samples. The English transcripts are first normalised and then converted with ARPABET phonetic transcription~\cite{jurafsky2019speech}. To avoid out-of-vocabulary (OOV) words, we update the lexicon dictionary with an English grapheme to phoneme (G2P) package~\cite{g2pE2019}. The Montreal forced aligner (MFA)~\cite{mcauliffe2017montreal} is used to obtain the phoneme duration.  Since, the audio frame-level sampling rate (16 ms) and the video sampling rate (30 fps for VOCAset and 25 fps for BIWI) are inconsistent, spline interpolation is applied in Stage II top layer output to align the network output with the video sampling rate.\looseness=-1

\noindent\textbf{Pipeline Architecture and Configuration.} The backbone structure in \cref{fig:overview} has some shared modules with FastSpeech-2~\cite{ren2020fastspeech}, i.e. the phoneme encoder, and decoder, that feature stacked self-attention layers. The variance adaptor is modified such that ground-truth phoneme duration is used for length regulation, and the style vector is added on top. The style encoder is initialised with two Wav2Vec2 networks, followed by stacked linear layers, \cref{fig:overview} (Right).

During Stage I training, we use the default train-validation-test split for ESD-EN, and split LJSpeech-2k with a 80-10-10 partitions. During Stage II training, both VOCAset and BIWI are split with 80-10-10 partitions. Adam optimizer with the same learning-rate (LR) scheduler in Vaswani~\etal\cite{vaswani2017attention} is used in Stage I, where the LR scheduler is modified for Stage II training with $lr=0.01$ and 1600 warm-up steps. Gradient-clip of 1.0 is applied.

Trained on a single A10G GPU, Stage I converged in about 24 hours after 500k steps. After stage I training, we show the T-SNE plot of the extracted style vectors over the ESD validation data, for different subjects and emotions in \cref{fig:stylevector}, where well-clustered patterns can be observed for both identities and emotions. We compare our extracted styles \cref{fig:stylevector}(a)+(b) with two baseline methods, the speaker embeddings in \cite{wan2020generalized} and the style embeddings in \cite{min2021meta}. 
 
In Stage I, the learned style vectors adapt to reconstruction of Mel-spectrum (audio speech). In Stage II, they adapt to generation of 3D vertices (visual speech). Despite audio speech and visual speech being highly correlated, there still exists some essential difference. E.g., two people whose voice sound similar may have very different articulatory movements and thus result in different facial motions. To cope with this discrepancy, the learned styles will evolve from audio-prominent attributes to visual-prominent attributes. To validate this idea, we randomly select six subjects (three males, three females) from VOCAset and BIWI, and extract the style vectors learned in Stage I and Stage II, respectively. Their TSNE plots are shown in \cref{fig:s1s2stylevector}. Take the VOCAset for example, the three males are distributed in the centre while the three females on the edge after Stage I, possibly due to the different range of pitches. However, after Stage II, M1 and F2 are pushed together, which might due to these two subjects having similar range of mouth openings.

\noindent
\textbf{Ablation on Laplacian-Mod Regularisation.} In Eq.~\ref{s2loss}, we use Laplacian-mod loss with $\lambda=1.0$ to mitigate high-frequency noise that occurs when extrapolating beyond the training set. Tikhonov regularization is often used at the deformation stage to avoid such artifacts, yet without considering the connections between vertices. We ablate Tikhonov loss weighted at 1.0 (\texttt{Tik1.0}) and 10.0 (\texttt{Tik10.0}) respectively, as well as when no regularisation is applied (\texttt{w/o}), and \texttt{groundtruth}. Without smoothing constraints (\texttt{w/o}), high-frequency artifacts can be generated. Obvious residual outliers persists for both levels of Tikhonov losses.

% https://docs.google.com/drawings/d/14da9DYlwhWXk7PRA0q05ymvoloCmPsvjsF7gKKR4Z4o/edit
\begin{figure}
  \begin{minipage}[c]{0.58\textwidth}
    \includegraphics[width=\textwidth]{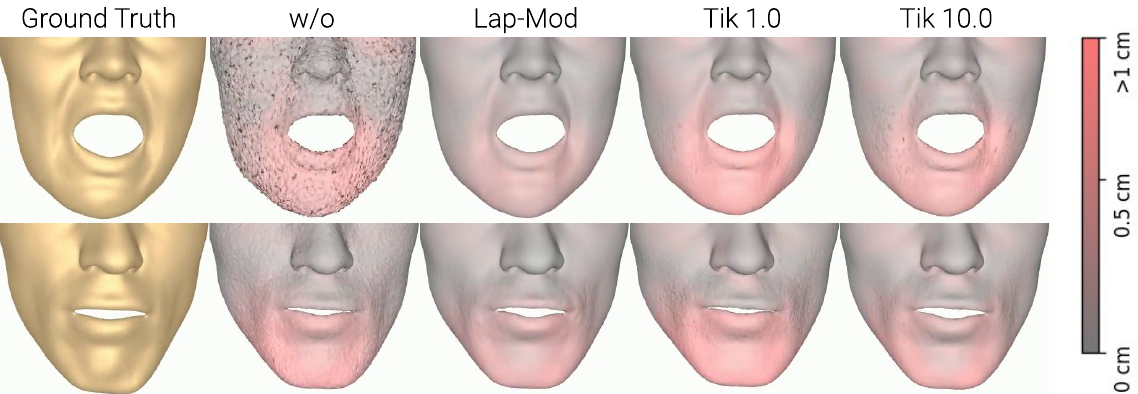}
  \end{minipage}\hfill
  \begin{minipage}[c]{0.40\textwidth}
    \caption{
       Ablation of the Laplacian-Mod (Lap-Mod) loss v.s. ground-truth, without (w/o) regularisation and with two levels of Tikhonov (Tik) regularisation. Red color represents errors.
    } \label{fig:alblaplacian}
  \end{minipage}
\end{figure}

% https://docs.google.com/drawings/d/1XVQ6yQbIvmAJmhcO-MQ7tFFspyRjcQdL0ele1cBb3Js/edit
\begin{figure*}
  \begin{minipage}[c]{0.25\textwidth}
    \caption{Mouth closures illustration, when the person is uttering  \textit{``our ex{\color{blue}p}eri{\color{blue}m}ent's {\color{blue}p}ositive outco{\color{blue}m}e''}. The highlighted bilabial phonemes relate to four mouth closures.}
    \label{fig:closures}
  \end{minipage}\hfill
  \begin{minipage}[c]{0.72\textwidth}
    \includegraphics[width=\textwidth]{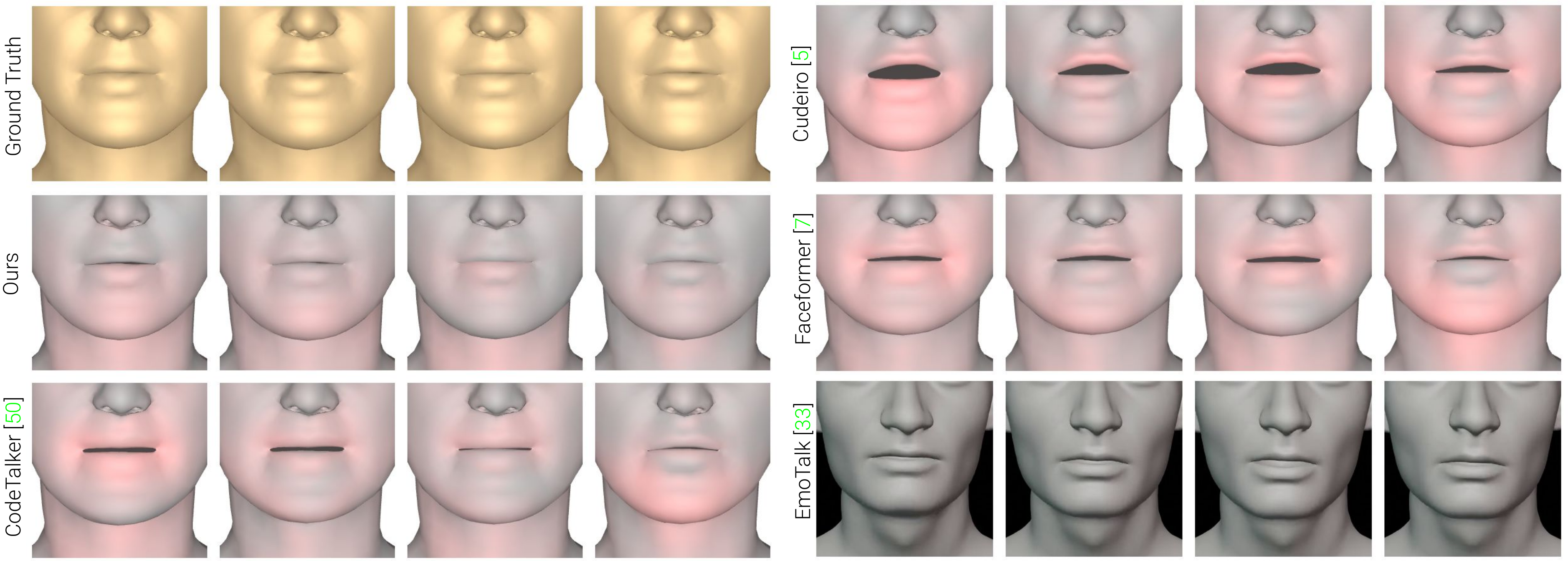}
  \end{minipage}
\end{figure*}

\noindent\textbf{Ablation on Different Loss Terms.}
Our loss terms contribute differently to the overall performance. We ablate over the following four situations: (1) When only vertex loss is applied without the weighting mask, denoted as \texttt{ver$\_$only}. (2) When weighted vertex loss is used, denoted as \textit{wei}\texttt{$\_$ver}. (3) When the weighting mask, vertex, landmark loss are all used, denoted as \textit{wei}\texttt{$\_$ver$\_$lmk}. (4) On top of \textit{wei}\texttt{$\_$ver$\_$lmk}, the Laplacian-mod loss is integrated, denoted as \texttt{Lap-Mod}. The following three quantitative evaluation metrics are employed: (a) vertex distance $d\_{ver}$ (b) landmark distance $d\_{lmk}$ (c) mouth closure distance with 3 frame latency allowance $d\_{close}_3$. The first two metrics are calculated as the per-vertex Frobenius norm. As for $d\_{close}_3$, the mouth closures are first detected, which is often related to the bilabial consonants ``B'', ``P'', ``M'', during which, $d\_{close}_3$ is calculated as the minimum estimated interior mouth distance within $\pm 3$ frames to allow for audio-visual off-sync latency. Results are averaged on the VOCA validation data.

% \begin{table}[H]
\begin{wraptable}{r}{7cm}
\vspace{-15px}
    \centering
    \caption{\textbf{Ablations over various loss terms.}}
    \begin{adjustbox}{width=\linewidth,right}
\begin{tabular}{rccc}
%\hline
 & $d\_{ver}$ & $d\_{lmk}$ & $d\_{close}_3$                 \\ \hline
\texttt{ver$\_$only}   & 0.062       & 0.201       & 0.052                           \\ %\hline
\textit{wei}\texttt{$\_$ver}        & 0.061       & 0.194       & 0.046                           \\ %\hline
\textit{wei}\texttt{$\_$ver$\_$lmk} & 0.061       & \textbf{0.183}       & \textbf{0.030} \\ %\hline
\texttt{Lap-Mod} & 0.061       & \textbf{0.185}       & \textbf{0.030} \\ \hline
\end{tabular}
  \label{tab:loss}
\end{adjustbox}
% \end{table}
  \vspace{-15px}
\end{wraptable}
% 
% 
% S2_from_scrach: Validation Step 20000, Vertices Loss: 0.0580, Vertices Dynnamics Loss: 0.0222,Landmark Loss: 0.1830, Landmark Dynamics Loss: 0.1211, Mouth Closure Loss: 0.0596, Laplacian Loss: 0.0000
Overall, $d\_{ver}$ and $d\_{lmk}$ stay consistent over the above ablation scenarios. When we apply the extra weighting mask, i.e. \textit{wei}\texttt{$\_$ver}, the mouth closure distance as well as the landmark distance improve, this proves that the proposed weighting mask prioritises the mouth region motions. Further, the applied mask encourages a quick convergence. Compared with \texttt{ver$\_$only}, only half the epochs (200 epochs vs 400 epochs) are needed before convergence, saving 50$\%$ training time. When we employ the landmark loss as well, i.e. \textit{wei}\texttt{$\_$ver$\_$lmk}, we achieve another approximate $35\%$ reduction on the mouth closure distance. With \texttt{Lap-Mod}, there is not much further change in above evaluation matrices. However, Laplacian-mod regularisation mitigates visually-noticeable high-frequency noise, as ablated earlier. %This validates the choice of our Stage II training target.

\noindent\textbf{Ablation of Stage II Training From Scratch}. Without Stage I training, we directly train Stage II from scratch. Comparing to the last row in Table \ref{tab:loss}, very similar vertex (0.058) and landmark (0.183) losses are obtained. Yet, there are 100$\%$ error increase in the mouth closure distance (0.030$\rightarrow$0.060), proving the essential contributions of disentangled embeddings. 

\subsection{Baseline Comparison}
\label{sec:baseline}
We compare our method with three state-of-the-art baseline methods, \textbf{Cudeiro~\etal~\cite{cudeiro2019capture}}, \textbf{Faceformer~\cite{fan2022faceformer}} and \textbf{CodeTalker~\cite{xing2023codetalker}} that are all trained on the VOCAset. We have also tested \textbf{EmoTalk~\cite{peng2023emotalk}}, whose in-house blendshapes are however not provided. Thus we cannot conduct a vertex-vertex comparison.  For a fair comparison, evaluations are applied onto the same test set that excludes the training data in all the baselines. All the baselines are conditioned on the same training identity. We use evaluation metrics of $d\_{ver}$, $d\_{lmk}$ and $d\_{close}$ with more strict 0, $\pm 1$ and $\pm 2$ frame off-sync allowance. Besides that, we extract the temporal variations $\sigma^2(\cdot)$ in a generated vertex sequence $\hat{\bf v}_t$, and compare the range of difference to that of ground-truth vertex $\bf v_t$, denoted as $d\_{\sigma} = \sqrt{|\sigma^2(\hat{\bf v}_t)-\sigma^2(\bf v_t)|}$. The maximal lip distance \cite{richard2021meshtalk} has also been calculated for lip synchronization evaluations, denoted as $lip $~(mm). We observe in Table \ref{tab:baselineloss} that, the range of temporal variations is similar for all methods, while our method shows consistent improvements over the vertex, landmark, maximal lip error, and mouth closure distance.

\vspace{-0.1cm}

\begin{table}[H]
\centering
\caption{\textbf{Quantitative evaluations with the baselines.}}
\begin{adjustbox}{width=0.8\linewidth,center}
\begin{tabular}{rcccccccl}
%\hline
\multicolumn{1}{c}{Method}                                                                                
& $d\_{ver}$     & $d\_{lmk}$     & $lip $~(mm) & $d\_{\sigma}$  & $d\_{close}_2$ & $d\_{close}_1$ & $d\_{close}_0$ & Condition \\ \hline
Cudeiro~\etal~\cite{cudeiro2019capture}       & 0.083          & 0.298          & 4.84        & 0.035          & 0.202        & 0.211        & 0.295          & \multirow{3}{*}{Fixed identity} \\ %\hline
Faceformer~\cite{fan2022faceformer}     & 0.082          & 0.258          & 4.27        & 0.034          & 0.113        & 0.145        & 0.246           \\ %\hline
CodeTalker~\cite{xing2023codetalker}     & 0.083          & 0.280          & 4.50        & 0.034          & 0.051        & 0.074        & 0.171           \\ \cline{2-9}
Cudeiro$^*$~\etal~\cite{cudeiro2019capture}   & 0.081          & 0.303          & 4.91        & 0.039          & 0.154        & 0.163        & 0.227          & \multirow{3}{2cm}{Optimal identity via greedy search} \\ %\hline
Faceformer$^*$~\cite{fan2022faceformer} & 0.081          & 0.264          & 4.30        & 0.034          & 0.106        & 0.137        & 0.227           \\ 
CodeTalker$^*$~\cite{xing2023codetalker} & 0.079          & 0.271          & 4.29        & 0.034          & \textbf{0.048}        & \textbf{0.067}        & 0.147           \\\hline
Ours                                          & \textbf{0.062} & \textbf{0.187} & \textbf{3.17}        & \textbf{0.032} & 0.070 & 0.075        & \textbf{0.139}  \\ \hline
\end{tabular}
\end{adjustbox}
\label{tab:baselineloss}
\end{table}

We further push the limits by conditioning the three baselines on all training subjects and choose the optimal one that yields closest-to-groundtruth vertices (minimum $d_{ver}$), and denote them as \textbf{Cudeiro$^*$~\etal~\cite{cudeiro2019capture}}, \textbf{Faceformer$^*$~\cite{fan2022faceformer}} and \textbf{CodeTalker$^*$~\cite{xing2023codetalker}}. After greedy search for the optimal condition, all the baselines show improvements in the mouth closure distance $d_{close}$. \textbf{Cudeiro$^*$~\etal~\cite{cudeiro2019capture}} and \textbf{Faceformer$^*$~\cite{fan2022faceformer}} remain consistent or slightly degraded in $d\_{lmk}$, $lip $~(mm) and $d\_{\sigma}$. For mouth closures, in the case of 1/2 off-sync allowance, \textbf{CodeTalker$^*$} provides better mouth closures. In the strict 0-latency scenario, our method shows advantages over the baselines, illustrated in \cref{fig:closures}.  %Mouth closures are perceptually important, which can be exploited for spoof detection~\cite{agarwal2020detecting}. 

Imitator~\cite{thambiraja2022imitator} is an alternative style-aware solution, which allows style adaption to new speakers given a training reference video, where an identity-specific motion decoding is finetuned from the training video. In ~\cite{thambiraja2022imitator}, two personalised models are provided on the VOCAset. Excluding the training samples for these two subjects, we run the above quantitative evaluations for our method and Imitator, in Tab \ref{tab:baselineloss2}. Imitator has obtained results close to but slightly worse than ours. This might be due to the rendered generation via Imitator still suffers some synchronicity issues, which results in large errors in geometric evaluations, especially for dynamic sequences. This can also be reflected by the huge difference in the mouth closure loss with difference off-sync latency. On the other hand, our model has explored an external aligner to provide explicit time alignment information, which mitigates the off-sync issues.
\begin{table}[H]
\centering
\caption{\textbf{Quantitative evaluations with Imitator~\cite{thambiraja2022imitator}.}}
\begin{adjustbox}{width=0.8\linewidth,center}
\begin{tabular}{rccccccc}
%\hline
\multicolumn{1}{c}{Method}                                                                                
& $d\_{ver}$     & $d\_{lmk}$     & $lip $~(mm) & $d\_{\sigma}$  & $d\_{close}_2$ & $d\_{close}_1$ & $d\_{close}_0$ \\ \hline
Imitator~\cite{thambiraja2022imitator}       & 0.079          & 0.190          & 3.15        & 0.033          & 0.110        & 0.144        & 0.262          \\ \hline
Ours                                          & \textbf{0.055} & \textbf{0.159} & \textbf{2.58}        & \textbf{0.029} & \textbf{0.070} & \textbf{0.077}        & \textbf{0.150}  \\ \hline
\end{tabular}
\end{adjustbox}
\label{tab:baselineloss2}
\end{table}

The above experiments are applied onto the well-built VOCAset and BIWI dataset. In order to test its robustness to out-of-domain data, we also run inference with some random in-the-wild speech. Our finding is, as long as accurate alignments are obtained via the external aligner, in-sync and fine-grained facial motions are generated. Finally, to better evaluate the overall quality, a user study has also been carried out. See supplementary materials for in-the-wild examples and details of the user study.%\looseness=-2

% \gb{We illustrated an example of mouth closure generation, where the two baselines cannot properly close the mouth.}
% https://docs.google.com/drawings/d/1aAe2WWamfiy7OQ4MQQfTWvFkPGd-e633UiU39-XnQd4/edit?usp=sharing
\begin{figure*}
  \centering
  \includegraphics[width=1.0\textwidth]{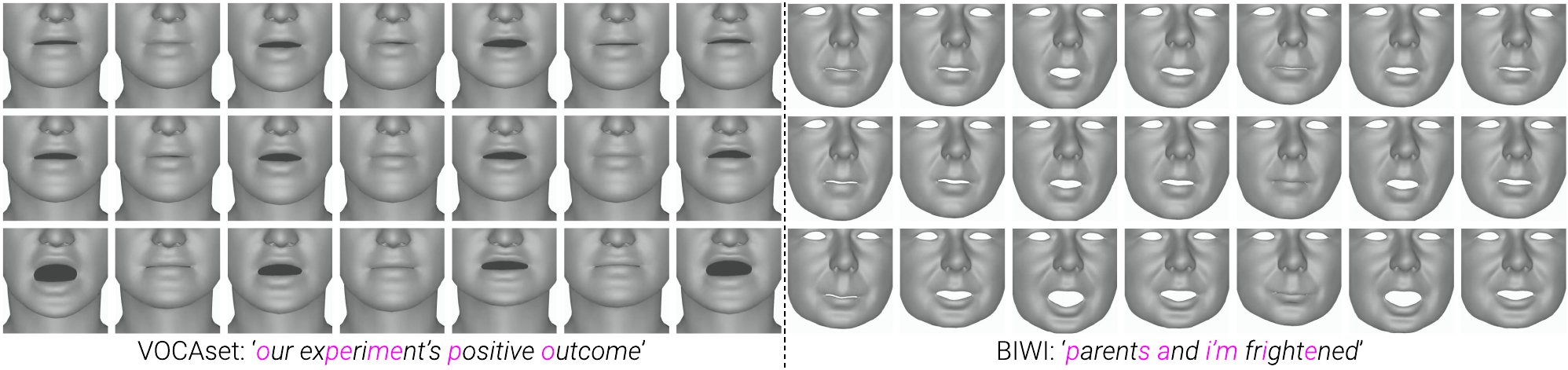}
    \caption{Two examples of extracting style from the driving audio (middle) and from a reference audio (bottom) v.s. the ground-truth (top), on VOCAset and BIWI datasets respectively. Highlighted letters are associated with different columns. The reference speaking styles tend to yield much larger mouth openings for these vowels.}
  \label{fig:change}
\end{figure*}

\subsection{Style- and Content-Aware Editing}
\label{sec:editing}
Due to the disentanglement of style and content components as well as the generative synthesis in our proposed pipeline, novel applications of style transfer and content editing are enabled, that were not possible in other audio-driven facial animation methods. 

\noindent\textbf{Style-Based Editing.} After Stage II fine-tuning, the pipeline is able to infer the implicit speaking style from the driving audio and control the facial animation generation. Besides the driving audio, the style embedding can be extracted from a reference audio of different speaking style, in order to change the style of the generated animation. For instance, \cref{fig:change} shows one example of directly using the driving audio and another reference audio for the style extraction, on both the VOCAset and BIWI datasets. We observe that direct inference generates close-to-groundtruth output, while style-transferred inference generates overall in-sync but different speaking styles.
% e.g. wider mouth openings. \gb{which fig? Besides that, from the second column and the forth column, the highlighted mouth pucker blendshape can be observed.}

\begin{figure*}
  \begin{minipage}[c]{0.35\textwidth}
    \caption{
       Modify the content transcript without directly modifying the audio signal. The original ``\textit{watch cartoons}'' (top) is replaced with ``\textit{read books}'' (middle) and muted (bottom) respectively. The muted segment smoothly transitions from the contextual word ``to'' to ``in'' during this short (0.85 sec) pause.
    } \label{fig:content}
  \end{minipage}\hfill
  \begin{minipage}[c]{0.62\textwidth}
    \includegraphics[width=\textwidth]{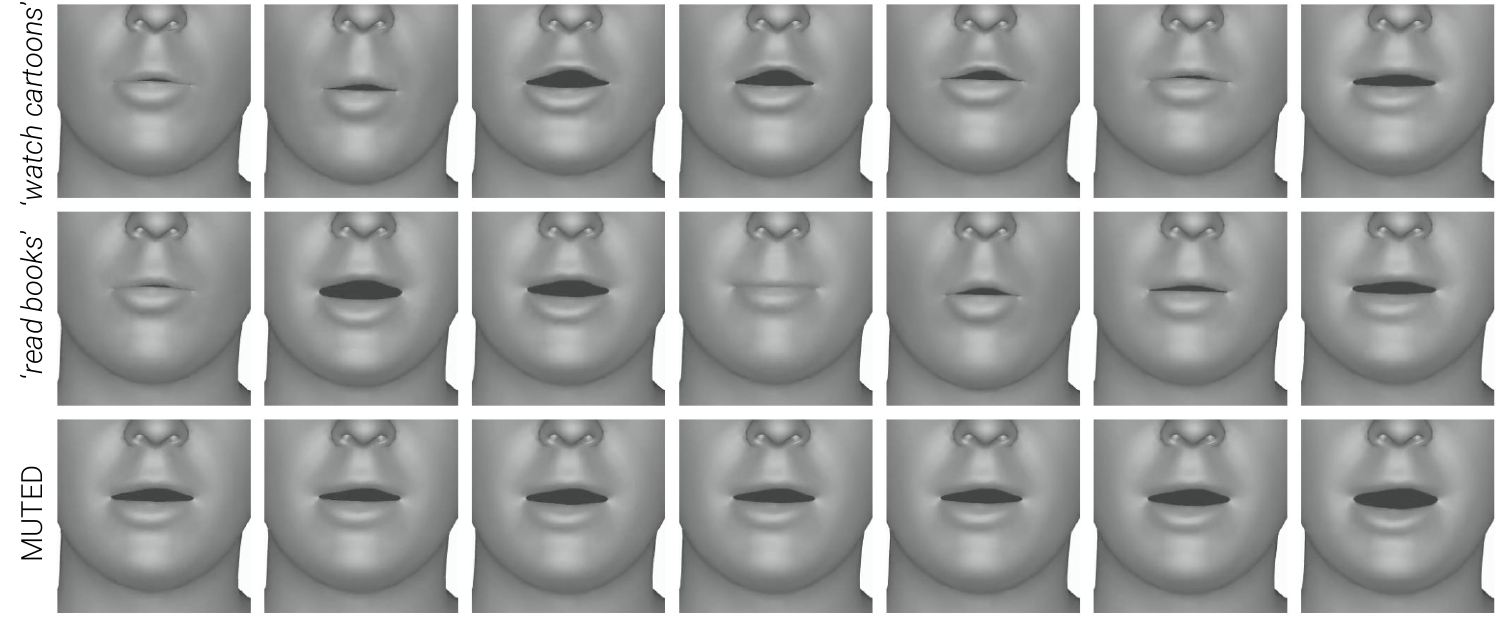}
  \end{minipage}
\end{figure*}

\noindent
\textbf{Content-Based Editing.} Besides style manipulation, we can also modify the aligned content and control the output. We choose an example where one person says \textit{``gregory and tom chose to \textbf{watch cartoons} in the afternoon''}, and we replace the words ``\textit{\textbf{watch cartoons}} (W AA1 CH K AA0 R T UW1 N Z)'' in the transcript with ``\textit{\textbf{read books}} (R IY1 D B UH1 K S)''. We also test muting ``\textit{\textbf{watch cartoons}}'' by replacing it with the silence token (sil) after the alignment. \cref{fig:content} samples the time snippet associated with the modified content. Using the original transcript (top), and two modified versions (bottom) yield different facial motions, accordingly. When we mute the selected segment, which is a short pause, there is a smooth transition process between the utterance of the two contextual words ``\textit{to}'' and ``\textit{in}'', rather than simply enforcing mouth closures.
% \vspace{-5px}

% \vspace{-0.25cm}

\noindent
\textbf{Computation analysis.}
The proposed method involves processing on both GPU (inference) and CPU (pre-processing). Tested on Nvidia GF RTX 3070 GPU, ignoring the overhead (load model, etc.), the average inference time is less than $0.1$ second for a 5-second-long audio, thanks to our efficient non-auto-regressive (NAR) architecture design, while the AR baselines (\cite{fan2022faceformer,xing2023codetalker}) take much longer, e.g. 2.5 seconds for \cite{fan2022faceformer}. However, our model's computational bottleneck is the pre-processing on a CPU machine, such as text normalisation, Mel-spectrum generation, and Montreal forced aligner (MFA). Particularly, MFA is the most time consuming (80+ seconds on an AMD Ryzen 9 5900HX processor) without batch-processing.

\section{Conclusion \& Limitation}
\label{sec:conclusion}

In this paper, we propose a two stage content and style aware audio-driven facial animation method. Style representation is achieved via speech-domain synthesis in Stage I, leveraging advances in TTS. In Stage II, the extracted style information contributes to style aware face animation generation and furthers adaption to the visual modality. Compared to other SOTA baselines, our method shows advantages over various evaluations, and provides more flexibility -- rather than assigning a specific identity style in the training data. Our method can infer the implicit style information from either the driving audio or a reference audio. Further, we can modify the content in the form of aligned text to generate new content articulations, while preserving the speaking style. We stress the essential limitation that our method relies an external audio-text alignment module, whose accuracy maybe affected by inaccurate transcripts as well as adverse acoustic environments, e.g. cross-talk, non-speech voice, and background noise. A potential solution to address these issues is to improve or augment the content representation, e.g. using frame-level semantic tokens of ``what'' is being uttered.

We also need to stress the potential negative ethical concerns. For instance, when combined with downstream tasks of audio-driven neural rendering and photo-realistic facial animation, misuse of this technology with malicious purposes cannot be overlooked. Proactive methods such as watermarking for spoofing detection should be considered.  

\bibliography{main}

\newpage
\appendix

\setcounter{figure}{0}
\renewcommand\thefigure{Appx.\arabic{figure}}
\setcounter{table}{0}
\renewcommand{\thetable}{Appx.\arabic{table}}

\section{User Study}
\begin{figure}[H]
\centering
  \centering
  \includegraphics[width=0.6\textwidth]{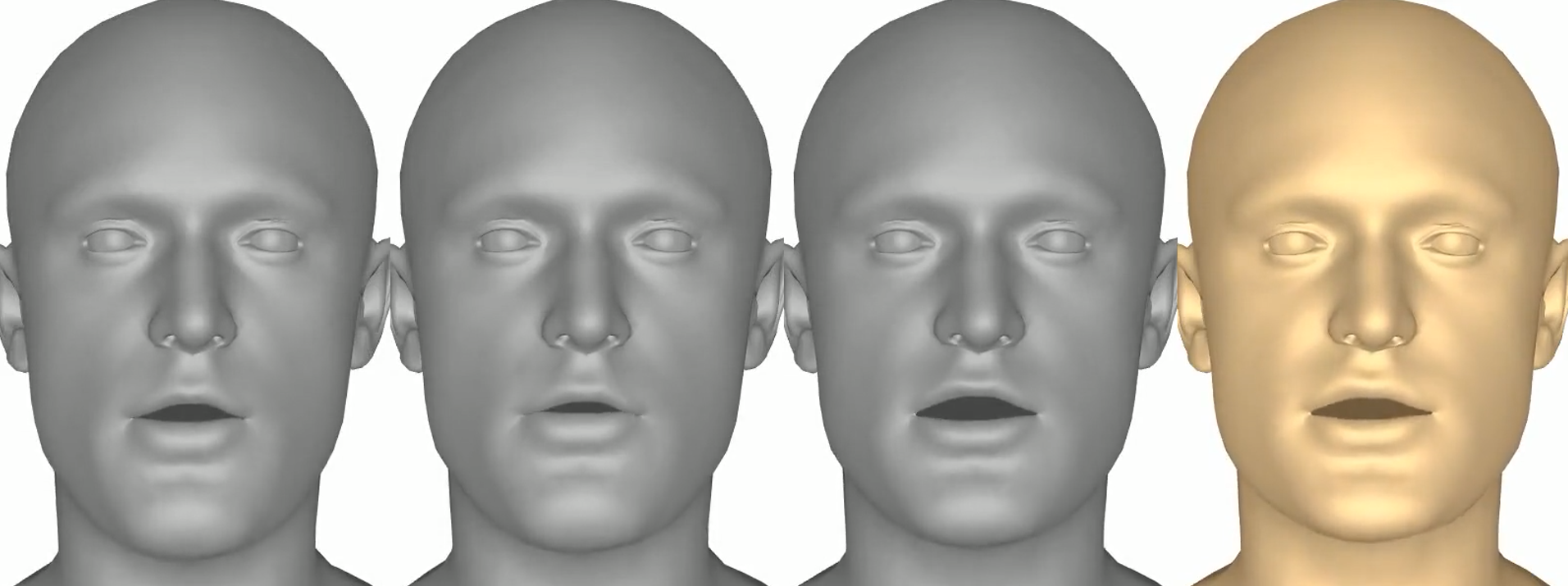}
\caption{One screenshot of our user study. Orders of the three compared methods are randomised (three left), to be compared against the ground-truth (highlighted far right).}
\label{fig:userstudy-head}
% \vspace{-10px}
\end{figure}

Besides the quantitative evaluations of our proposed method with  state-of-the-art baseline methods in the main submission, we have also carried out a user study to perceptually evaluate the quality of the animations, with two baselines \textbf{Cudeiro~\etal~[5]} and \textbf{Faceformer~[7]}. We have 12 participants, and each was given 10 randomly (not cherry-picked) chosen sets of samples. In each set, results from our method and the two baselines, as well as the ground-truth signal are presented at the same time. We have highlighted the ground-truth on the far right with a different color for comparisons. To avoid bias, orders of the three compared methods in each set are also randomised. \cref{fig:userstudy-head} shows one screenshot of the user study. 

\begin{figure}
\centering
  \centering
  \includegraphics[width=0.9\textwidth]{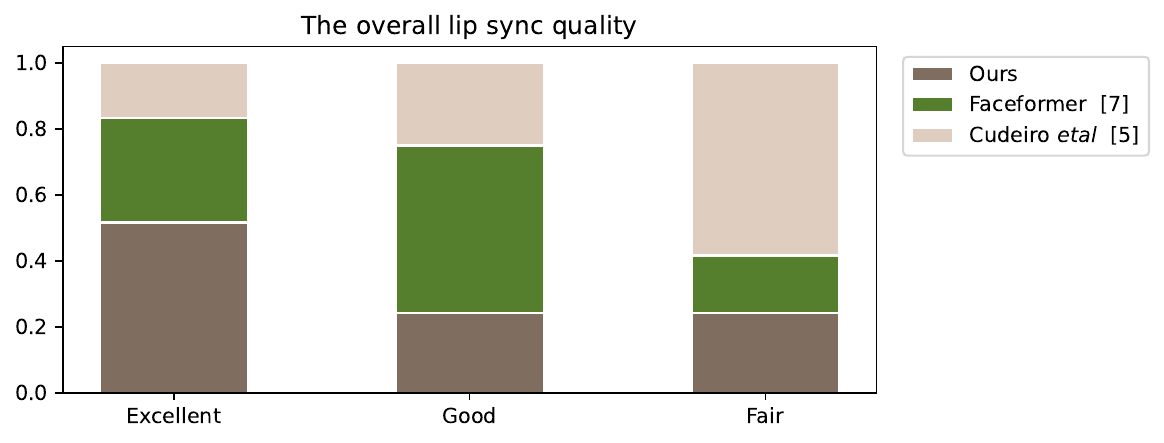}
\caption{Subjective evaluations of the three methods in terms of overall lip sync quality.}
\label{fig:userstudy}

\end{figure}

The participants are asked to evaluate the overall lip sync quality with three scores [Excellent, Good, Fair] as compared to the ground-truth, and the scoring is performed by a forced ranking method. Many factors can contribute to individual participant's judgement, e.g. articulation accuracy, audio-visual latency. Participants are required to zoom in their web browsers and watch the clips multiple times before scoring, and suggested to manually change the playback speed of each clip to, e.g., x0.5. The average scores are then calculated for all tests, see \cref{fig:userstudy}. It can be observed that our proposed method outperforms the two baseline methods, with half of the ranked excellent samples being chosen from our method. This advantage is probably introduced by the extra inferred style information. The Faceformer method~[7] comes the second, where its majority cases are evaluated as good. Cudeiro~\etal~[5], on the other hand, is mostly considered as fair, whose degradation might be caused by the lack of long contextual information. 

\section{Dataset and Pipeline Architecture}

\begin{figure}[h]
  \centering
  \includegraphics[width=\linewidth]{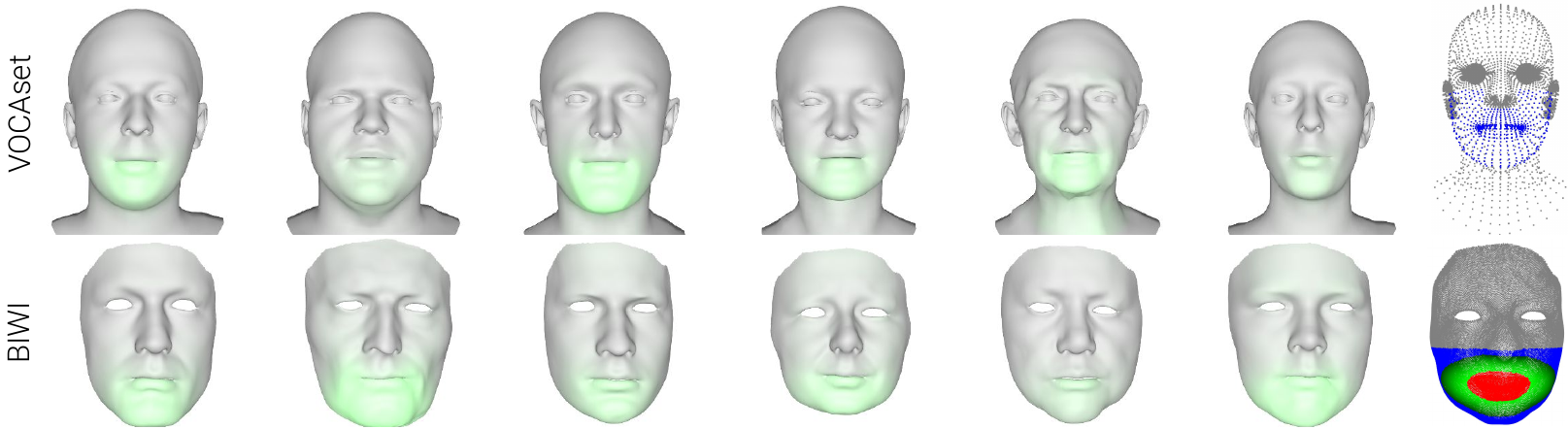}
  \caption{The VOCAset topology (top) and the BIWI topology (bottom), where the intensity of the shaded green color denotes the motion variations of each vertex, which most happens in the mouth and jaw regions. The VOCAset is more stable, with less movements over the face boundaries. Based on the distributions of the facial motions, we have empirically applied two different vertex masks (Sec 3.2) for Stage II training, with two segments for VOCAset and 4 segments for BIWI respectively highlighting different colors (far right), to prioritise audio-correlated facial motions.}
  \label{fig:identity}
  % \vspace{-10px}
\end{figure}

\textbf{Dataset.} We have considered both the VOCAset and BIWI datasets in our stage II training, that contain registered high quality audio/4D mesh pairs. The VOCAset head topology includes 5023 points that are un-evenly distributed, covering the whole head and neck. On the other hand, the BIWI topology contains 23770 vertices that are relatively even but high-densely distributed, and it covers only the front narrow face region. For each dataset, we randomly chose three males and three females, and calculated the temporal variations per vertex for each subject. The two head typologies are shown in \cref{fig:identity}. The VOCAset and BIWI vertex values are two orders of different. As a result, we have scaled the VOCAset by 100 thus consistent training parameters can be used for both datasets. 

When calculating the geometric vertex loss, a weighting mask has been applied to prioritise vertices more related to audio-correlated facial motions. The empirically-set weighting masks for the VOCAset and BIWI are shown in the right column of \cref{fig:identity} where different colors denote different mask values. The two mask values in VOCAset are $[0.5, 1.5]$, and the four mask values in BIWI are $[0.5, 1.0, 2.0, 4.0]$.  

% https://docs.google.com/drawings/d/1y_rFi5-RWIPWxEBgAr0EfEZ10BrlbSnR0KffpXMkcog/edit?usp=sharing

\textbf{Pipeline Architecture.}
Our model's backbone structure contains a phoneme encoder that represents content, a style encoder with disentangled style (prosody, identity, intonation .etc) representation. A parameterless variance adaptor combines the style and length-regulated content, which are then fed into a decoder. The phoneme encoder and the decoder employ the same structure as these modules being used in FastSpeech2, i.e. stacked transformer layers. The style encoder features two pre-trained Wav2Vec2 sub-networks, which are pre-trained on large datasets (LibriSpeech, Voxceleb~1, IEMOCAP), in total 1000+ hrs. This usage of high-resource audio data has encouraged the extracted style embedding to be disentangled from content. Two project layers ${\cal F}_{s1}$ and ${\cal F}_{s2}$ are applied onto the decoder output, to generate either Mel-spectrum (Stage I) or head meshes (Stage II). Depending on the audio-mesh datasets that have different number of vertices $N_v$ (e.g., 5123 for VOCAset, and 23770 for BIWI), the last layer in ${\cal F}_{s2}$ changes with different head typologies. 

Details of the parameter setting in our model are listed in Table~\ref{tab:moduleparams}. 

\begin{table}[H]
    \centering
    \caption{\textbf{Module configurations in the pipeline.}}
\begin{tabular}{rl}
%\hline
{Module} & Setup     \\ \hline
Phoneme Encoder       & 4$\times$ Transformer FFTBlock                    \\ %\hline
       & Head=2, HiddenSize=256, FilterSize=1024\\ %\hline
       & KernelSize=[9, 1], Dropout(0.2)                    \\ %\hline
Style Encoder     & 2 $\times$ Wav2Vec2.0, Concat\\ %\hline
       & MaxPool1d(KernelSize=2, Stride=2) \\ %\hline  
       & Linear(64) + ReLU + Dropout(0.2) \\ %\hline       
       & Linear(256) + Tanh \\ %\hline
Decoder       & 6$\times$ Transformer FFTBlock                    \\ %\hline
       & Head=2, HiddenSize=256, FilterSize=1024\\ %\hline
       & KernelSize=[9, 1], Dropout(0.2)                    \\ 
${\cal F}_{s1}$    & Linear(80) \\
${\cal F}_{s2}$    & Linear(100) + Tanh + Linear($3N_v$)\\
\hline
\end{tabular}
  \label{tab:moduleparams}
\end{table}

\section{More Results}
\begin{figure}
\centering
  \centering
  \includegraphics[width=\textwidth]{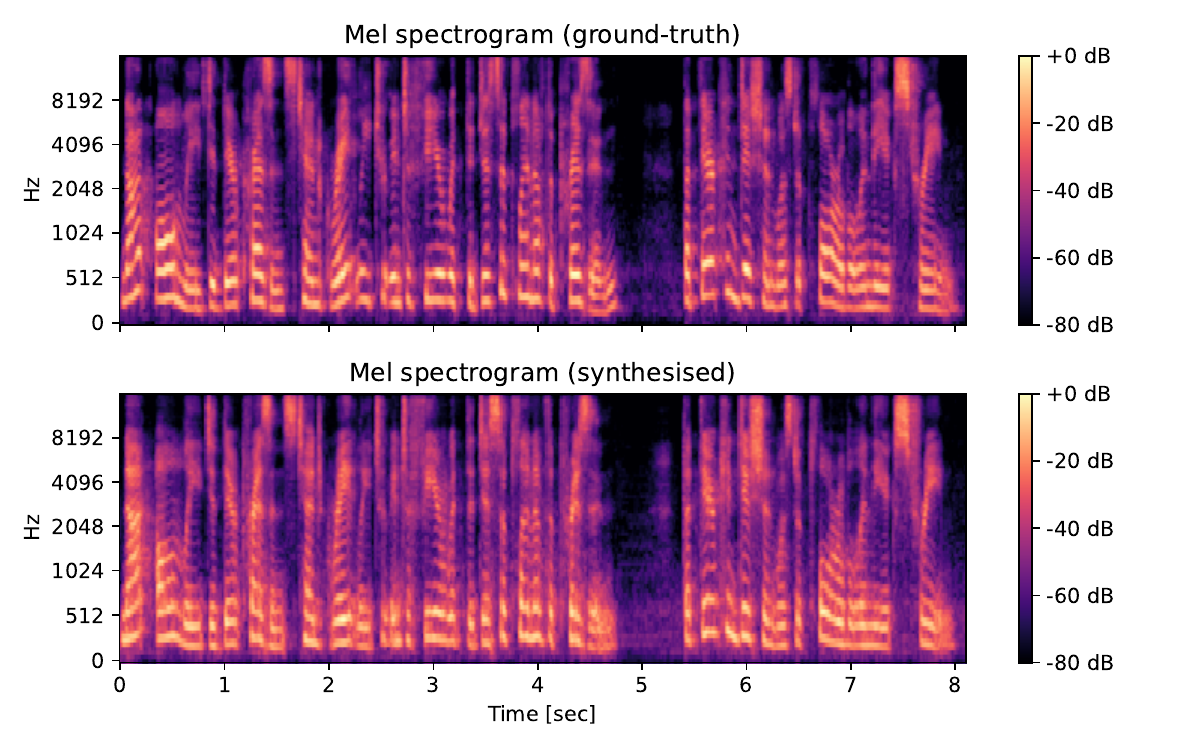}
\caption{One example of the ground-truth (top) and synthesised (bottom) speech Mel-spectrum after Stage I training on the LJSpeech-2K dataset. }
\label{fig:s1training}
\vspace{-10px}
\end{figure}

\textbf{Stage I.} Our method is trained with speech synthesis task of Mel-spectrum self-reconstruction in Stage I, using two datasets: ESD-EN and LJSpeech-2K. \cref{fig:s1training} shows the model is capable of synthesizing speech signals close to the ground-truth. The ESD-EN dataset has overall more blurred spectrum glimpses, while LJSpeech-2K shows a clear harmonics pattern. The synthesised speech, controlled by the extracted style information, has resembled these features. 

\begin{figure}
\centering
  \centering
  \includegraphics[width=\textwidth]{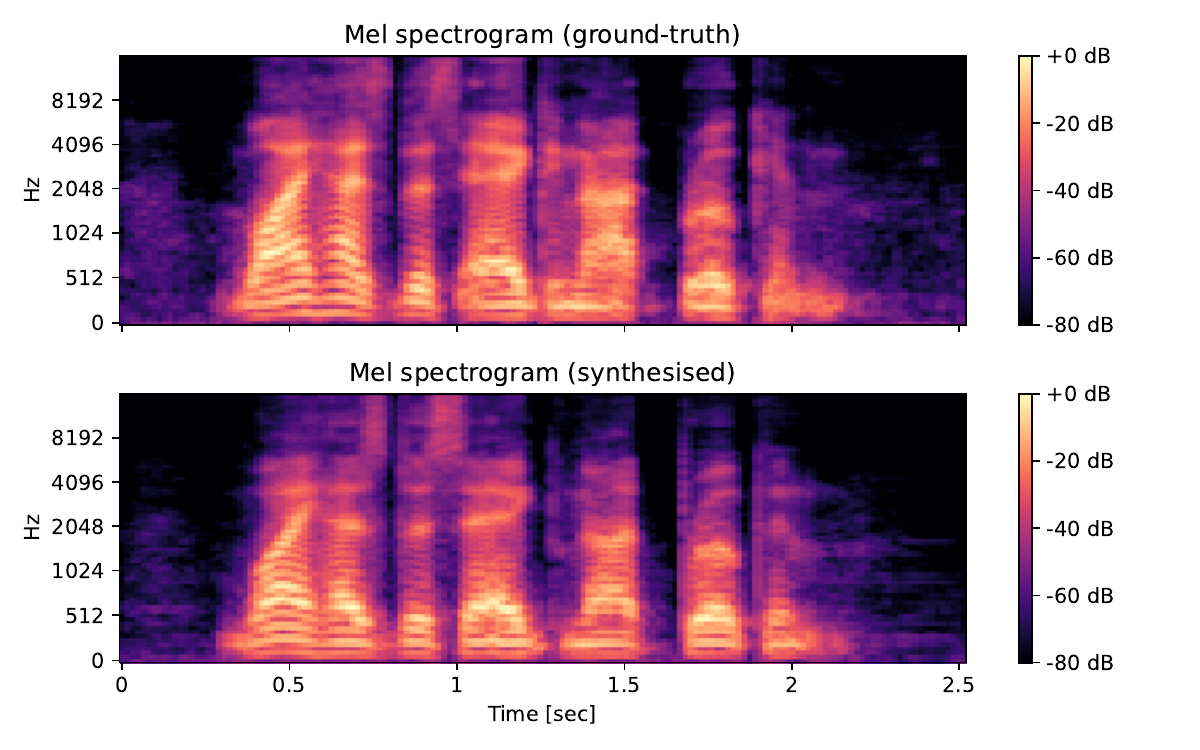}
\caption{One example of the ground-truth (top) and synthesised (bottom) speech Mel-spectrum after Stage I training on the ESD-EN dataset.}
\label{fig:0020_000017}
\vspace{-10px}
\end{figure}

\newpage
\textbf{Stage II.} Depending on the employed head topology (5023 vertices for VOCAset and 23370 vertices for BIWI), the last linear layer of ${\cal F}_{s2}$ is different. We have trained two models for these two datasets. Besides the ablation of different loss terms on VOCAset reported in the main manuscript, we also did the similar study on the BIWI dataset. Consistent improvements are observed when gradually vertex loss, weigting mask, landmark loss are added. However, overall larger distances are yielded for BIWI compared to VOCAset. This is because the BIWI dataset is more challenging with more spread facial motions beyond the mouth/jaw region, which is also consistent with the large temporal variations we reported earlier. 

\begin{table}[H]
% \vspace{-15px}
    \centering
    \caption{\textbf{Ablations over various loss terms on BIWI.}}
\begin{tabular}{rccc}
%\hline
                                            & $d\_{ver}$ & $d\_{lmk}$ & $d\_{close}_3$                 \\ \hline
\texttt{ver$\_$only}   & 0.14       & 0.26       & 0.15                           \\ %\hline
\textit{wei}\texttt{$\_$ver}        & 0.15       & 0.25       & 0.11                           \\ %\hline
\textit{wei}\texttt{$\_$ver$\_$lmk} & 0.14       & 0.24       & \textbf{0.09} \\ \hline
\end{tabular}
  \label{tab:lossbiwi}
  % \vspace{-15px}
\end{table}

\begin{figure}
\centering
  \centering
  \includegraphics[width=0.8\textwidth]{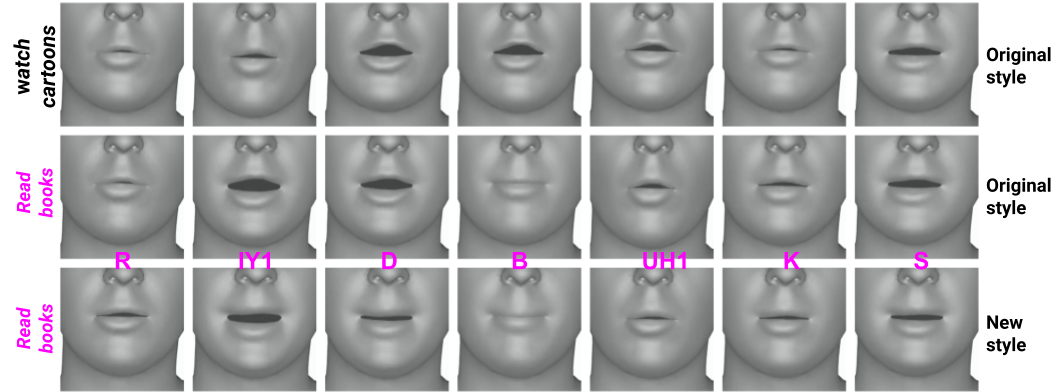}
\caption{The segment when ``watch cartoons'' (top) is directly changed to ``read books'' (middle), and using a new reference style (bottom). With the new style, the facial motions for phonemes R, IY1, D and S are different.}
\label{fig:contentandstyle}
\end{figure}

We have shown some examples of style transfer and content modification isolatedly in the main manuscript. These two modifications can be achieved simultaneously, \cref{fig:contentandstyle} shows one example of changing ``watch cartoons'' to ``read books (R IY1 D B UH1 K S)'' using the original style, and a new reference style respectively. To the best of our knowledge, none of any its counterpart methods can achieve simultaneous style transfer and content modification applications.

% https://docs.google.com/drawings/d/1JxE-5a-TsTJS72A34K0DT3m6yLcZZztCV682qrPzw40/edit

\end{document}